\documentclass[aps,prc,reprint,superscriptaddress,nofootinbib,amsmath,amssymb,longbibliography]{revtex4-2}
\usepackage{adjustbox}
\usepackage[colorlinks]{hyperref} 
\usepackage[normalem]{ulem}

\usepackage{booktabs}
\usepackage{siunitx}
\usepackage{multirow}
\usepackage{amsmath}
\usepackage{dblfloatfix}
\usepackage{placeins}
\usepackage{bm}
\usepackage{upgreek}
\usepackage{physics}
\usepackage{makecell}

\begin{document}

\hypersetup{
	colorlinks=true,
	linkcolor=black,
	citecolor=black,
	urlcolor=black
}

\title{Polarized Electron Scattering from Light Nuclei at High Energies}

\author{Minh Truong Vo}
\email{Corresponding author: Minh Truong Vo; Institute of Fundamental and Applied Sciences, Duy Tan University, Ho Chi Minh City, 70000, Vietnam; Faculty of Natural Sciences, Duy Tan University, Da Nang, 50000, Vietnam; vominhtruong@duytan.edu.vn}
\affiliation{Institute of Fundamental and Applied Sciences, Duy Tan University, Ho Chi Minh City 70000, Vietnam}
\affiliation{Faculty of Natural Sciences, Duy Tan University, Da Nang City 50000, Vietnam}

\author{Vu Dong Tran}
\affiliation{Institute of Fundamental and Applied Sciences, Duy Tan University, Ho Chi Minh City 70000, Vietnam}
\affiliation{Faculty of Natural Sciences, Duy Tan University, Da Nang City 50000, Vietnam}

\author{Quang Hung Nguyen}
\email{Co-corresponding author: Quang Hung Nguyen; Institute of Fundamental and Applied Sciences, Duy Tan University, Ho Chi Minh City, 70000, Vietnam; Faculty of Natural Sciences, Duy Tan University, Da Nang, 50000, Vietnam; nguyenquanghung5@duytan.edu.vn}
\affiliation{Institute of Fundamental and Applied Sciences, Duy Tan University, Ho Chi Minh City 70000, Vietnam}
\affiliation{Faculty of Natural Sciences, Duy Tan University, Da Nang City 50000, Vietnam}

\date{\today}

\begin{abstract}  
\vspace{0.5em}
We present a theoretical approach to investigate the scattering of polarized electrons from light nuclei using the multipole expansion for the scattering cross section within the framework of the unified electroweak theory. Scattering processes corresponding to different electron polarizations are analyzed and compared with the unpolarized electron scattering investigated earlier. Besides, the contribution of both polarized and unpolarized terms to the scattering cross section is examined. Numerical calculations for stable $^{6,7}$Li and unstable $^7$Be nuclei using the Weinberg-Salam model show that the longitudinal polarization and weak interaction are not explicitly correlated when electrons scatter at $ \theta \simeq 0^{\circ}$ across all energy scales. A strong correlation emerges at the other scattering angles when the electron energy exceeds 10 GeV. This study provides additional information about nuclear structure and uncovers the role of electron polarization and its correlation with the weak interaction in each process, thus offering a more complete picture of electron-nucleus scattering.

\end{abstract}

\maketitle
%%%%%%%%%%%%%%%%%%%%%%%%%%%%%%%%%%%%%%%%%%%%%%%%%

{\Large \textbf{1. Introduction}}
\vspace{5pt}

Electron polarization plays an important role in studying nuclear structure by means of scattering. Technical advancements in generating polarized electron beams \cite{moor08} enable us to test the predictions from theoretical models and obtain additional data regarding their scattering on both unoriented and oriented nuclei \cite{moor08,weig64,dapo18,bemi23}. In general, the polarization of an electron beam can either remain unchanged or be altered after scattering with a target \cite{donn86,dapo18, manaut05, jakubassa18, dahiri22}. This means that the scattering cross section and asymmetry depend on both the magnitude of electron polarization and how it changes \cite{donn86,lipp89,adhi22,bacc24,gonza15}. One must either calculate the polarization density matrix through the scattering cross section or precisely produce the polarized beams for experimental use \cite{weig64,moor08,abbo16,srom99}. Scattering processes at high energies are often of particular interest since electrons have only longitudinal polarization along the direction of momentum transfer \cite{donn86}.

\vspace {5pt}
It was initiated long ago that the multipole expansion
for the electromagnetic cross section can be applied to
examine the asymmetry resulting from electron polarization
and nuclear spin orientation, and clarify their impact \cite{weig64,donn86}. A similar expansion for the electroweak cross section, which includes the contribution of weak interaction, has also been proposed \cite{keri84,luon03,luon23,truo24}. While the nuclear form factors in prior studies \cite{weig64,donn86,keri84,luon03} were either approximated or treated as parameters, they were calculated in a precise way in later works \cite{luon23,truo24} using the unified electroweak theory combined with the
many-particle shell model and fractional parentage coefficients \cite{jahn51,elli53,will63}. In particular, by employing the V-A structure of the neutral weak current along with appropriately chosen harmonic oscillator parameters, the latest
study in Ref. \cite{truo24} has successfully described the experimental data for the unpolarized electron scattering on $^{6,7}$Li and $^{7}$Be nuclei at MeV energies \cite{rand66,suel67}, and provided predictive insights at GeV energies. 

\vspace {5pt}
To date, a comprehensive study of high-energy polarized electron-nucleus scattering with respect to the changeabilities of electron polarization in electroweak interaction processes, which could reveal information beyond what is inaccessible through pure electromagnetic interaction processes and unpolarized electron sources, has not been performed. This paper presents an investigation of all possible longitudinally polarized electron scatterings
from unoriented light nuclei using
the multipole expansion within the framework of the unified electroweak theory. It also provides a basis for future studies on the asymmetry resulting solely from the electron polarization and weak interaction, since nuclear spin orientation is excluded. Accordingly, the polarized scattering cross sections will be calculated and compared with the unpolarized scattering cross sections via their ratios. On the one hand, this enables us to compensate for the experimental limitations. Indeed, we can either perform the experiments for unpolarized or polarized electron-nucleus scattering and then deduce the result for the other case using the investigated ratios. On the other hand, the above ratios also help to clarify the contribution of electron polarization in different scattering processes when the weak interaction is involved. In addition, the impact of electron polarization and its correlation with the weak interaction are also examined through the ratios between the polarized and unpolarized terms in each scattering process.

In addition to the Introduction, the general formalism of the multipole expansion for the electroweak scattering cross section of polarized electrons on unoriented light nuclei will be briefly presented in Section 2. However, the given calculations only apply to inclusive scattering processes, and the nucleon knockout mechanism will not be taken into account. In Section 3, we apply the formalism of the multipole expansion presented in the previous section to several specific cases. More specifically, they include the elastic electron scattering processes on $^6$Li, $^7$Li, and $^7$Be nuclei when these nuclei are in their ground state. Accordingly, the scattering cross section for the possible electron polarization states before and after scattering will be considered separately and compared with that for unpolarized electrons. Analytical and numerical calculations related to the scattering cross section and the nuclear form factors can be found in the Appendix. In the Summary and Outlook section, we summarize some key results and outline prospects for the following research.

\vspace{10pt}

{\Large \textbf{2. Formalism}}
\vspace{5pt}

The scattering process of high-energy polarized electrons on light nuclei can be described within the framework of the unified electroweak theory using the first-order Born approximation. Accordingly, the interaction between the electron and the nucleus, which includes electromagnetic and weak neutral current interactions, is mediated through the exchange of a virtual photon $\gamma$ and an intermediate boson $Z^{\circ}$. The differential scattering cross section is written as \cite{luon23,truo24} 

\begin{equation} 
	  \sigma = \frac{4m_e^2\varepsilon'}{f\varepsilon } \sum_{if}^\text{---} \left|M_{fi} \right|^2,
\end{equation}
with $m_e$ being the electron mass, $\varepsilon$ and $\varepsilon'$ being the electron energies before and after scattering, and $f$ being the nuclear recoil factor \cite{donn86,will63}. The notation $ \sum\limits_{if}^\text{---} $ denotes averaging over the initial spin states and summing over the final spin states for both the electron and nucleus. The scattering amplitude $ M_{fi} $ is the sum of two parts corresponding to the electromagnetic interaction and the weak interaction of the form
\begin{equation}
    M_{fi} =-\frac{4\pi}{Q^2} [\alpha\overline{u}'\gamma_\mu u J_F^\mu (Q)+ \lambda \overline{u}'\gamma_\mu (g_V + g_A\gamma_5)uJ_Z^\mu (Q) ].
\end{equation} 

The notation $ Q = K - K' =(\omega, \textbf{q}) $ is the 4-dimensional momentum transfer with $ K = (\varepsilon, \textbf{k}) $ and $ K'=(\varepsilon', \textbf{k}') $ being the electron momenta. $ u = u(K, S) $ and $ u' = u' (K', S') $ are the spinor state amplitudes of the electron before and after scattering, respectively. $S$ and $S'$ are the vectors of electron polarization, which can be expressed approximately through the momenta by $ S^\mu \approx \xi K^\mu/m_e $ and $ S'^\mu \approx \xi' K'^\mu/m_e $ with $\xi$ and $\xi'$ being the electron polarizations or helicities, whose values are equal to +1 or $-1$ \cite{donn86,luon23}. $\alpha$ and $g$ are the electromagnetic and weak coupling constants. $\lambda = - G_{F} m_{Z}^2 Q^2 / [2 \sqrt{2} \pi (m_{Z}^2 - Q^2)],  G_{F} = g^2/(4 \sqrt{2} m_{Z}^2 \cos^2 \theta_{\text{W}}) $ with $\theta_{\text{W}}$ being the Weinberg angle, and $m_{Z}$ being the $Z^{0}$ boson mass. $g_{V} = -1/2 + 2 x_{\text{W}},  g_{A} = -1/2$ and $x_{\text{W}} \equiv  \sin^2 \theta_{\text{W}}$ are the weak interaction parameters in the Weinberg-Salam model. $J_F^\mu(Q)$ and $J_Z^\mu(Q)$ are the electromagnetic and weak currents of the nucleus, $\mu = 0,1,2,3$.

\vspace {5pt}
For brevity, when performing analytical calculations, we set $\hbar = c = 1 $, and their values will be restored to ensure the correct dimensions of related physical quantities in the numerical evaluations. Substitute the expression (2) into (1) and perform the summation over electron states and then calculate the matrix traces, the scattering cross section depends on the contraction products in terms of $X_{\mu \nu}H_{F/FZ/Z}^{\mu \nu} $ and $ Y_{\mu \nu}H_{F/FZ/Z}^{\mu \nu}$. More specifically, $X_{\mu \nu} = K_\mu K'_\nu + K'_\mu K_\nu + g_{\mu \nu}Q^2/2$ and $Y_{\mu \nu}  = -i \varepsilon_{\mu \nu \alpha \beta} K^\alpha K'^\beta$ are the symmetric and asymmetric electron tensors respectively with $ g_{00} = +1, g_{11} = g_{22} = g_{33} = -1 $, and $\varepsilon_{0123} = + 1$. The nuclear tensor $ H_F^{\mu \nu} = \sum\limits_{if}^\text{---} J_F^{\mu*}J_F^\nu$ corresponds to the pure electromagnetic interaction, $H_{FZ}^{\mu \nu} = \frac{1} {2} \sum\limits_{if}^\text{---} (J_F^{\mu*}J_Z^\nu + J_Z^{\mu*}J_F^\nu)$ corresponds to the interference of the electromagnetic and weak interactions, and $ H_Z^{\mu \nu} = \sum\limits_{if}^\text{---} J_Z^{\mu*}J_Z^\nu$ belongs to the pure weak interaction. The summation and average are now restricted to the nuclear states. 

We now choose a cyclic coordinate system with the unit vectors as $\bm{\upzeta}_0 = \textbf{e}_z$, $\bm{\upzeta}_{\pm1} = - (\pm\textbf{e}_x +i\textbf{e}_y)/\sqrt{2}$, where three unit vectors in the Cartesian coordinate system are determined by $\textbf{e}_z = \underline{\textbf{q}}$, $\textbf{e}_y = \underline{\textbf{k}\times\textbf{k}'}$, $\textbf{e}_x = \textbf{e}_y\times\textbf{e}_z$. Accordingly, the above contraction products can be further computed by expanding the nuclear transition currents in terms of the components with defined angular momenta, which are called the nuclear multipole form factors.
They provide information about the structure of the nucleus along with scattering processes.

\vspace {0pt}
Recall that the electromagnetic current is conserved, while the longitudinal component of the weak neutral current is very small and can be neglected compared to the remaining components. As a result, applying the Wigner-Eckart theorem for the nuclear transition currents, we can obtain the expression of the scattering cross section in the form \cite{luon23, truo24}
\begin{equation} 
	  \sigma_{\text{total}}^{\text{pol.}} (\xi,\xi') = \sigma_{\text{total}}^{0} + \sigma_{\text{total}}^{\xi \xi'}.
\end{equation}
For more details on the calculation steps, please see Section A in the Appendix. The first term in the expression (3), which does not involve electron polarization, is a sum of
\vspace{1pt}
\begin{align}
	\sigma_{F}^{0} &= \frac{\alpha^{2} \varepsilon'}{f \varepsilon Q^{4}} A_{1}, \tag{4a} \\
    \sigma_{FZ}^{0} &= \frac{2 \alpha \lambda \varepsilon'}{f \varepsilon Q^{4}} (g_{V} B_{1} + g_{A} B_{2}), \tag{4b}	\\
     \sigma_{Z}^{0} &= \frac{ \lambda^{2} \varepsilon'}{f \varepsilon Q^{4}} \left[ \left(g_{V}^{2} + g_{A}^{2}\right) C_{1} + 2 g_{V} g_{A} C_{2} \right] \tag{4c},	
\end{align} 
while the second term is a respective sum of
\begin{align}
	\sigma_{F}^{\xi \xi'} &=  \frac{\alpha^{2} \varepsilon'}{f \varepsilon Q^{4}} [ \xi \xi' A_{1} +(\xi + \xi') A_{2} ],	\tag{5a} \\
	\sigma_{FZ}^{\xi \xi'} &= \frac{2 \alpha \lambda \varepsilon'}{f \varepsilon Q^{4}} \{ [g_{V} \xi \xi' + g_{A} (\xi + \xi')] B_{1} \nonumber \\
	 & \phantom{000000.} + [g_{V}(\xi + \xi') + g_{A} \xi \xi'] B_{2}  \},	\tag{5b}	\\
	\sigma_{Z}^{\xi \xi'} &= \frac{ \lambda^{2} \varepsilon'}{f \varepsilon Q^{4}} \{ [(g_{V}^2 + g_{A}^2) \xi \xi' + 2 g_{V} g_{A} (\xi + \xi')] C_{1} \nonumber \\
	&\phantom{000000} + [(g_{V}^2 + g_{A}^2)(\xi + \xi') + 2 g_{V} g_{A} \xi \xi'] C_{2} \}. \tag{5c}
\end{align} 

The coefficients $A_{1}, A_{2}, B_{1}, B_{2}, C_{1}$, and $C_{2}$ are determined in terms of the multipole form factors as follows
\begin{align}
	A_1 &= H \sum_L \left\{ 
    u_C (F_L^C)^2 
    + u_T \left[ (F_L^E)^2 + (F_L^M)^2 \right] 
    \right\},	\tag{6a} \\
	A_2 &= 0,	\tag{6b}	\\
	B_1 &= H \sum_L \left[ 
    u_C F_L^C V_L^C 
    + u_T ( F_L^E V_L^E + F_L^M V_L^M) \right], \tag{6c}   \\
    B_2 &= H \sum_L u_T'\left( F_L^E A_L^M + F_L^M A_L^E \right), \tag{6d}   \\
    C_1 &= H \sum_L \{ 
    u_C \left[ (V_L^C)^2 + (A_L^C)^2 \right] + u_T \left[ (V_L^E)^2 \right. \notag\\
    &\left. + (V_L^M)^2 + (A_L^E)^2 + (A_L^M)^2 \right]\}, \tag{6e}   \\
    C_2 &= 2 H \sum_L u'_T \left( V_L^E A_L^M + V_L^M A_L^E \right), \tag{6f}
\end{align}
where $ H \equiv 4\pi/(2J + 1)$ with $J$ being the total spin of the nucleus in the initial state, and $\abs{J-J'} \le L \le {J+J'}$ according to the selection rule. $ u_C = 2\varepsilon^2(1-x^2)$, $u_T = \varepsilon^2(1 + x^2)$, and $u'_T = 2\varepsilon^2x $ are the kinematic coefficients, $x \equiv \text{sin}(\theta/2) $ with $\theta$ being the scattering angle. The symbol $F$ denotes the electromagnetic form factor, while
$V$ and $A$ represent the vector and axial form factors of the weak interaction. The superscripts $C$, $E$, and $M$ of the form factors imply their Coulomb, electric, and magnetic components, respectively.

\vspace{5pt}
The expressions (3), (4a)-(4c) and (5a)-(5c) with the coefficients given by (6a)-(6f) are the complete multipole expansion for the high-energy polarized electron scattering cross section on light nuclei within the framework of the unified electroweak theory. The above result will return to that of the electromagnetic scattering case described by Weigert-Rose \cite{weig64} when the weak interaction is not considered, as well as to that given by Willey \cite{will63} when the electron polarization is not taken into account. Numerical calculations for the case of unpolarized electron scattering cross sections \cite{truo24} using the nuclear shell model with the harmonic oscillator parameter $\beta $ = 1.1526 \text {fm}$^{-2} $ for $ ^6\text{Li} $ nucleus and with $\beta$ = 1.3055 \text {fm}$^{-2}$ for $^7\text{Li}$ and $^7\text{Be}$ nuclei describe the experimental data in Refs. \cite{rand66,suel67} rather well at MeV energies. The results obtained also provide more information about the scattering processes at higher energies. Particularly, an application to the $^7$Li nucleus allows us to distinguish the contribution of the elastic scattering process at the ground state and that of the quasielastic scattering process in the excitation from the ground state to the nearest state, whereas one can only obtain the combined data of these two processes in experiments.

\vspace{5pt}
However, polarized electron scattering processes have not yet been considered. So, we now assign $\xi\,(\xi') = +1$ for the right-handed polarization and $\xi\,(\xi') = -1$ for the left-handed polarization. We will consider all possible cases of electron polarization as follows: Both incoming and outgoing electron beams are unpolarized. This case has been studied in detail in our recent study \cite{truo24}. Next, electron polarizations before and after scattering are preserved, i.e., $\xi =\xi'$. Then, an initially polarized beam becomes unpolarized, i.e., $\xi = +1/-1$ and $\xi'=0$ (or $\xi'$ being taken over +1 and $-1$). This case will yield the same result as an initially unpolarized electron beam becomes polarized after scattering, since $\xi$ and $\xi'$ play equivalent roles in expressions (5a)-(5c). Finally, an electron beam reverses its initial polarization after scattering, i.e., $\xi = -\xi'$.

\vspace{10pt}

{\Large \textbf {3. Results and Discussions}}

\vspace{5pt}
It can be seen that the scattering cross section (3) will be equivalent to that of unpolarized electron scattering when we take the summation over all possible electron polarization states. This implies that an unpolarized electron beam has the spin components uniformly distributed in all directions \cite{dapo18}. 

In addition, with $\xi = \xi' = +1 / -1$, the contribution of pure electromagnetic interaction via the electron polarization term (5a) to the total differential cross section (3) is identical to that via the unpolarized term (4a) due to the coefficient $A_2 = 0 $, a consequence of the multipole expansion for the nuclear transition currents. Moreover, $\sigma_{F}^{11} = \sigma_{F}^{-1-1} = \sigma_{F}^{0}$ also shows no difference between left-polarized electrons and right-polarized electrons. This relates to the fact that the pure electromagnetic interaction preserves electron polarization as previously mentioned in Ref. \cite{donn86} and does not cause parity violation \cite{gonza15}. Meanwhile, the contributions of the polarized terms involving the weak interaction (5b)–(5c) to the total cross section are unequal and different from those of the respective unpolarized terms (4b)–(4c). 

In particular, when $\xi = 0$ and $\xi' = +1/-1$ (or $\xi = +1/-1$ and $\xi' = 0$), only the last two terms (5b) and (5c) make a nonzero contribution due to $\sigma_{FZ}^{0\pm 1}= \sigma_{FZ}^{\pm 10} \ne 0$ and $ \sigma_{Z}^{0 \pm 1} = \sigma_{Z}^{\pm10} \ne 0 $ while $\sigma_{F}^{0\pm1} = \sigma_{F}^{\pm 10} =0$. It means that, for the cases under consideration, only the weak interaction makes electron polarization affect the scattering processes, while the electromagnetic interaction does not, regardless of whether or not the weak interaction is present. As a result, with an initially unpolarized electron beam ($\xi = 0$), we can obtain a scattered beam that is either unpolarized ($\xi' = 0$) or left-/right-polarized ($\xi' = +1/-1$) with different probabilities. In contrast, suppose we could create an initial electron beam that is completely left-/right-polarized ($\xi = +1/-1$), the multipole expansion of the scattering cross section shows that we could obtain either a scattered beam with the polarization unchanged or an unpolarized beam after scattering ($\xi' = 0$), which have nonzero probabilities. In general, the two possible processes ($\xi = 0$, $\xi' = +1/-1$) and ($\xi = +1/-1$, $\xi' = 0$) arise only when the weak interaction is present and it affects the scattering processes through the terms $\sigma_{FZ}^{0\pm 1}, \sigma_{Z}^{0 \pm 1}$, and $\sigma_{FZ}^{\pm 10}, \sigma_{Z}^{\pm10}$. Thus, the weak interaction may be a cause that makes an initially polarized electron beam become unpolarized after scattering, and vice versa. However, this needs to be verified experimentally through numerical calculations for specific nuclei, which will be presented in the next section below. 

And finally, the total differential cross section (3) will vanish when the electron polarizations before and after scattering have opposite signs. This implies that the electroweak interaction cannot cause the spin orientation of all electrons to be reversed. In other words, the spin-flip scattering cross section, which is determined by formula (3), is zero. This leads to a degree of polarization in our case for the electroweak interaction $P = 1$, which matches the result pointed out in Refs. \cite{manaut05, dahiri22} when electrons scatter at relativistic energies via only the purely electromagnetic interaction. Compared with the aforementioned studies, our result differs notably in that it accounts for weak interactions, along with the processes ($\xi = 0$, $\xi' = +1/-1$) and ($\xi = +1/-1$, $\xi' = 0$), as just analyzed above. The specific expressions corresponding to all cases under discussion can be readily derived from the formulas (5a)–(5c). 

\vspace{5pt}
The above analysis highlights common features of high-energy scattering processes of polarized electrons on light nuclei, elucidates the role of electron polarization and its correlation with the electromagnetic and weak interactions. Obviously, this information cannot be obtained through the multipole expansion of the scattering cross section using the electromagnetic theory or by considering unpolarized electron scattering processes. To get more detailed results, we need to calculate all the multipole form factors present in the expressions (4a)-(4c) and (5a)-(5c) of the scattering cross section (3). It can be seen that the dependence of the two components (involving and not involving electron polarization) on the multipole form factors is the same through the coefficients (6a)-(6f). However, the possible multipole components to be calculated, a finite number according to the selection rule, are different when considering specific state transitions or various nuclei. 

Since the multipole form factors are the reduced matrix elements of the respective operators, we first need to construct the explicit expressions for these operators, along with the formulas for calculating their matrix elements. While the calculation of single-particle matrix elements is quite simple, the calculation of matrix elements for multiparticle systems is much more complicated. In principle, we can convert multiparticle matrix elements into single-particle matrix elements for further calculation via the particle density matrix \cite{donn79} or using the fractional parentage coefficient \cite{luon23,truo24,will63}. In particular, using the formulas for calculating the reduced matrix elements given in Refs. \cite{luon23,truo24}, which are derived from the angular momentum theory combined with the fractional parentage coefficient method and nuclear shell model, all the multipole form factors in (6a)-(6f) can be computed directly without requiring any additional approximations (please see Sections B and C in the Appendix). Calculations for the scattering of unpolarized electrons on $^6$Li and $^7$Li nuclei have shown good agreement with experimental data using the harmonic oscillator parameters aforementioned in the previous section. Furthermore, the convection currents in these cases are also retained. So, in the following, we will consider the scattering processes of polarized electrons on $^6$Li, $^{7} \text{Li}$ and $^{7} \text{Be}$ nuclei, in a similar way as performed for the unpolarized electron cases, since the latter results have been experimentally verified. 

\vspace{5pt}
We limit our considerations here to only the elastic scattering processes when these nuclei are in their ground states in order to clarify the roles of electron polarization and the weak interaction, as well as their correlation when comparing equivalent processes. Accordingly, while the form factors of the $^6$Li nucleus  are $F_0^C$, $F_2^C$, $F_1^M$, $V_0^C$, $V_2^C$, $V_1^M$, $A_1^C$ and $A_1^E$, the two $^7$Li and $^7$Be nuclei have more than form factors including $F_3^M$, $V_3^M$, $A_3^C$ and $A_3^E$. In general, they all depend on the nucleon form factors and the harmonic oscillator parameter. We will use the nuclear form factors, which were calculated directly, given in Ref. \cite{truo24} for numerical calculations below, where the harmonic oscillator parameter $\beta $ = 1.1526 \text {fm}$^{-2} $ for $ ^6\text{Li} $ nucleus and the value $\beta$ = 1.3055 \text {fm}$^{-2}$ for $^7\text{Li}$ along with $^7\text{Be}$ while the nucleon form factors are parameterized as dipoles as described in Refs. \cite{luon23,truo24,wang06,povh08,chang18}. 

\vspace{5pt}
Now we rewrite the nuclear form factor expressions for elastic scattering from $ ^6\text{Li} $ as follows
\vspace{0pt}
\begin{align}
	F_0^{C}(q) &= \frac{\sqrt{3}}{\sqrt{\pi}}\, G_E^{s}\, J_{0},	\tag{7a} \\
	F_2^{C}(q) &= 0,	\tag{7b}	\\
	F_1^{M}(q) &= -\,\frac{q}{\sqrt{\pi}\,m_N}\, G_M^{s}\, J_{0}, \tag{7c}   \\
    A_1^{C}(q) & = \,\frac{\sqrt{2}\,q}{\sqrt{\pi}\,m_N}\,\beta_A^{(0)}\,G_A^{s}\,J_{0}, \tag{7d}   \\
    A_{1}^{E}(q) &= \frac{2}{\sqrt{\pi}} \, \beta_{A}^{(0)} \, G_{A}^{s} \, J_{0}, \tag{7e}   \
\end{align}
and for elastic scattering from $ ^7\text{Li} $ and $^7\text{Be} $ as follows
\vspace{0pt}
\begin{align}
	F^{C}_{0}(q) &= \frac{1}{\sqrt{\pi}} \, \tilde{X}_{F} \, J_{0},	\tag{8a} \\
	F^{C}_{2}(q) &= \frac{3}{5\sqrt{\pi}} \, \tilde{X}^{F} \, J_{2},	\tag{8b}	\\
	F^{M}_{1}(q)& =
    -\frac{\sqrt{5}\, q}{9\sqrt{2\pi}\, m_{N}} \left[ \tilde{X}^{F}(J_{0}+J_{2})
    + 3\tilde{Y}^{F}\!\left(J_{0}-\frac{3}{25}J_{2}\right)
    \right], \tag{8c}   \\
    F^{M}_{3}(q) &=
    -\frac{3\sqrt{3}\, q}{5\sqrt{5\pi}\, m_{N}}
    \, \tilde{Y}^{F} \, J_{2}, \tag{8d}   \\
    A^{C}_{1}(q)& =
    \frac{\sqrt{5}\, q}{3\sqrt{\pi}\, m_{N}}
    \, \tilde{Y}^{A}
    \left(
    J_{0} + \frac{6}{25} J_{2}
    \right), \tag{8e} \\
    A^{C}_{3}(q)& =
    \frac{9\, q}{5\sqrt{5\pi}\, m_{N}}
    \, \tilde{Y}^{A} \, J_{2}, \tag{8f} \\
    A^{E}_{1}(q) &=
    \frac{\sqrt{10}}{3\sqrt{\pi}}
    \, \tilde{Y}^{A}
    \left(
    J_{0} - \frac{3}{25} J_{2}
    \right), \tag{8g} \\
    A^{E}_{3}(q)&=
    \frac{6\sqrt{3}}{5\sqrt{5\pi}}\,
    \tilde{Y}^{A}\, J_{2}. \tag{8h}   \
\end{align}

For more detailed calculations, see Section D in the Appendix. In the above expressions, $m_N$ is the nucleon mass, while $\tilde{X}^{F} = 3G^{s}_{E} \mp G^{v}_{E}$, $\tilde{Y}^{F} = G^{s}_{M} \pm G^{v}_{M} $ and $\tilde{Y}^{A}
=
\beta_{A}^{(0)} G^{s}_{A}
\pm
\beta_{A}^{(1)} G^{v}_{A} $ with the upper sign for $ ^7\text{Li} $ and the lower sign for $^7\text{Be} $. The radial matrix elements are determined by $J_{0}=(1-2z/3)e^{-z}$ and $J_{2}=2ze^{-z}/3$ with $z={q^{2}}/({4\beta})$. The notations $\beta_{A}^{(0)}$ and $\beta_{A}^{(1)}$ are the parameters of the Weinberg–Salam model. The isoscalar ($s$) and isovector ($v$) components are determined through the form factors of the proton and neutron by $G^{s}_{E}=({G^{p}_{E}+G^{n}_{E}})/{2}$, $G^{v}_{E}=({G^{p}_{E}-G^{n}_{E}})/{2}$, $G^{s}_{M}=({G^{p}_{M}+G^{n}_{M}})/{2}$, $G^{v}_{M}=({G^{p}_{M}-G^{n}_{M}})/{2}$, $G^{s}_{A}=({G^{p}_{A}+G^{n}_{A}})/{2}\equiv G_P$, $G^{v}_{A}=({G^{p}_{A}-G^{n}_{A}})/{2}\equiv G_A$ \cite{luon23}. The vector form factors are derived from the respective electromagnetic form factors following the procedure described in Ref. \cite{luon23}.

\begin{figure}
	\includegraphics[width=0.5\textwidth]{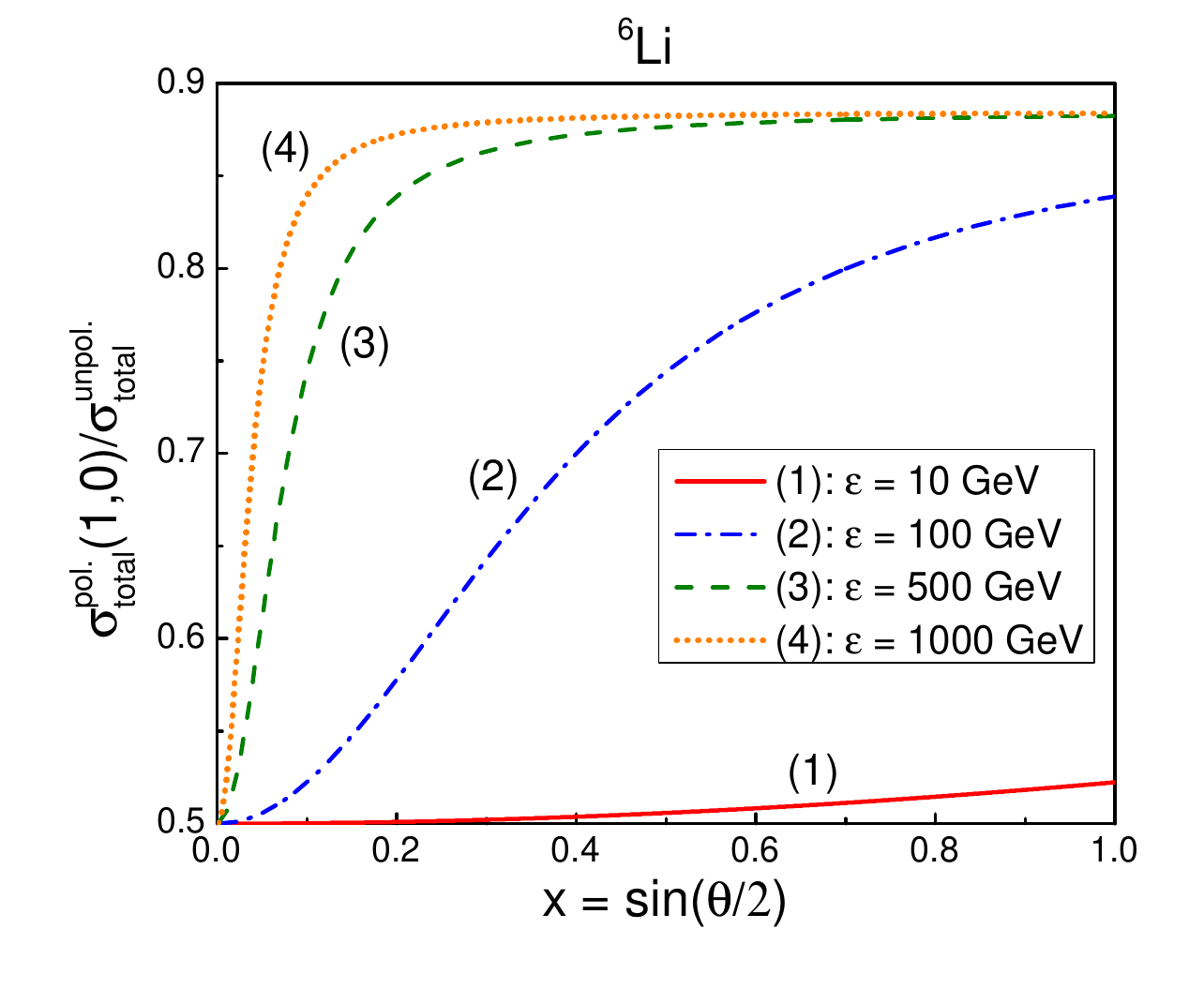}
    \vspace{-28pt} 
	\caption{%	
		The polarized cross sections $\sigma_{\text{total}}^{\text{pol.}}(1,0)$ versus the unpolarized ones $\sigma_{\text{total}}^{\text{unpol.}}$ obtained for $^{6}$Li at different electron energies $\varepsilon$ = 10, 100, 500, and 1000 GeV and scattering angles $\theta$.}
	\label{fig1}
	\vspace{-15pt}
\end{figure}

\begin{figure}[htbp]
	\centering
	\includegraphics[width=0.5\textwidth]{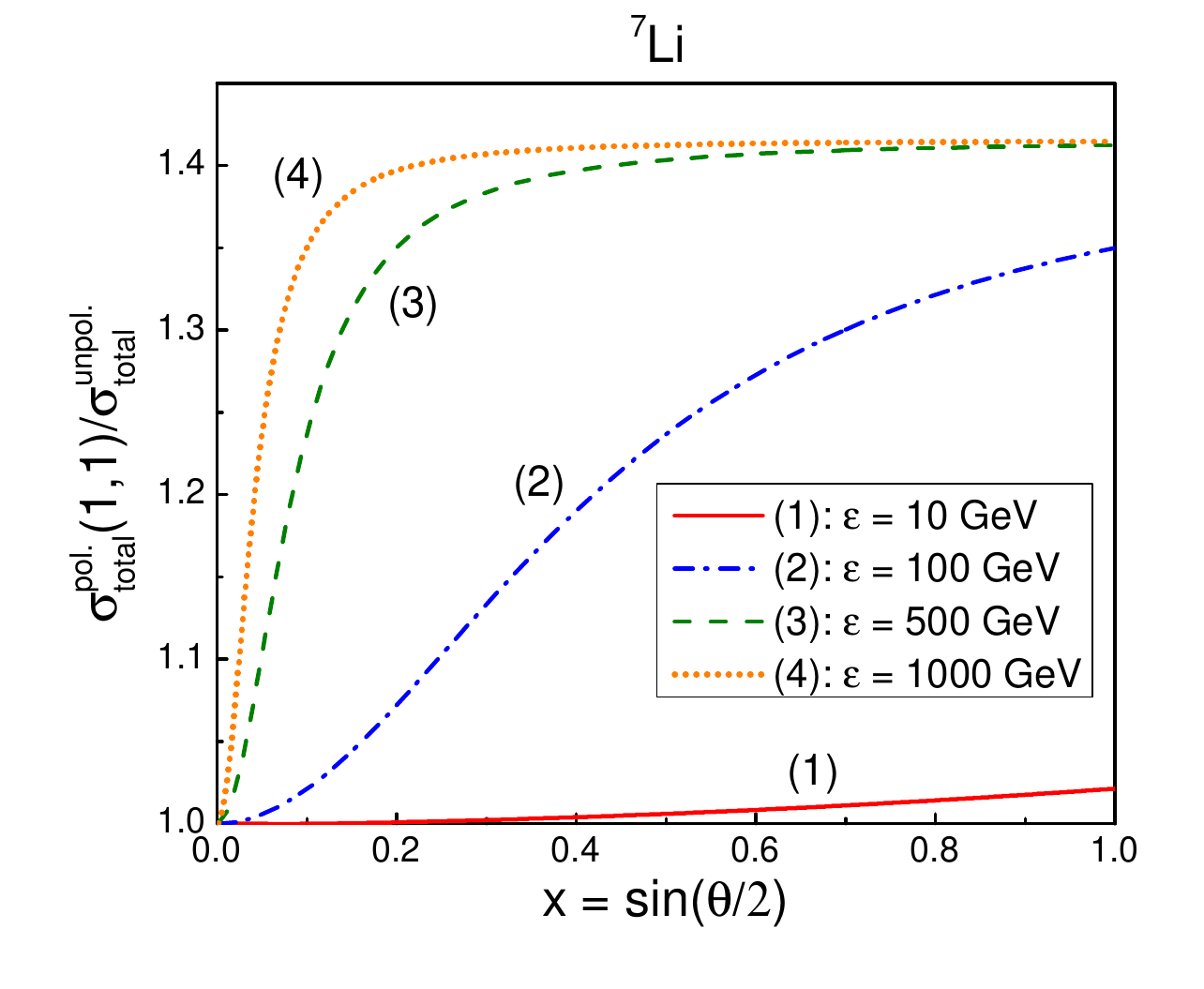}
	\vspace{-28pt}
	\caption{%
		The polarized cross sections $\sigma_{\text{total}}^{\text{pol.}}(1,1)$ versus the unpolarized ones $\sigma_{\text{total}}^{\text{unpol.}}$ obtained for $^{7}$Li at different electron energies $\varepsilon$ = 10, 100, 500, and 1000 GeV and scattering angles $\theta$.}
	\label{fig2}
	\vspace{-15pt}
\end{figure}
\vspace {5pt}
Now, we investigate two typical polarized scattering cross sections, $\sigma_{\text{total}}^{\text{pol.}}(1,0)$ and $\sigma_{\text{total}}^{\text{pol.}}(1,1)$, and compare them to the unpolarized one $\sigma_{\text{total}}^{\text{unpol.}}$, with $\sigma_{\text{total}}^{\text{unpol.}} = 2 \sigma_{\text{total}}^{0}$. Here, $\sigma_{\text{total}}^{\text{pol.}}(1,0)$ represents that the incoming right-handed polarization electron beam becomes unpolarized after scattering. Similarly, $\sigma_{\text{total}}^{\text{pol.}}(1,1)$ implies both incoming and outgoing electron beams being the right-handed
polarization ones. Numerical calculations are performed with various incident electron energies and scattering angles. The MATLAB source codes used for these calculations are provided as supplementary material. The relevant parameters are also given in Section E of the Appendix. 

\vspace {5pt}
Fig. 1 shows the calculated $\sigma_{\text{total}}^{\text{pol.}}(1,0)/\sigma_{\text{total}}^{\text{unpol.}}$ ratio for $^6$Li, whereas Fig. 2 plots the computed $\sigma_{\text{total}}^{\text{pol.}}(1,1)/\sigma_{\text{total}}^{\text{unpol.}}$ one for $^7$Li. The $\sigma_{\text{total}}^{11}/\sigma_{\text{total}}^{0}$ ratio for $^7$Be is also displayed in Fig. 3. All the cross section ratios for the three nuclei under consideration are listed in Table I. It can be seen that at energies not exceeding $10$ GeV, $\sigma_{\text{total}}^{\text{pol.}}(1,0)$ is just approximately half of $\sigma_{\text{total}}^{\text{unpol.}}$, while $\sigma_{\text{total}}^{\text{pol.}}(1,1)$ is almost equal to the latter. Notably, the values of the above ratios are nearly independent of the scattering angle and remain unchanged over the MeV energy range. This means that from the experimental data of the unpolarized electron scattering cross section on light nuclei, e.g., data in Tables I and II in Ref. \cite{suel67}, we can deduce the values of the polarized electron scattering cross section at the respective energy and scattering angle. In contrast, if the polarized electron scattering cross section at an arbitrary energy and scattering angle is measured, we can also deduce the scattering cross section for the unpolarized case that might not have been available earlier. This gives us a more complete picture of both unpolarized and polarized electron scattering cross sections and compensates for experimental limitations. That the ratios under consideration are almost constant could be attributed to the negligible contributions
of the weak interaction in the polarized terms (5b)-(5c) at
such an energy range \cite{luon23}. Meanwhile, when the electron energy passes 10 GeV, the above ratios strongly depend on the energy and scattering angle, except at 0$^\circ$. It can be explained by the fact that at such an energy scale, the weak interaction emerges and correlates with electron polarization to make polarized cross sections considerably increase, as seen in Figs. 1–3. This feature depicts the difference between pure electromagnetic scattering and electroweak scattering at high energies. 

\vspace {5pt}
A slight difference is observed when comparing the cross section ratios for three nuclei. For $^6$Li, since its multipole form factors have only isoscalar components, the given ratios are almost independent of the nuclear form factor when using the Weinberg-Salam model, with the parameter $\beta_A^{(0)} = 0$. This feature is similar to the asymmetry (which is also determined by the scattering cross section ratio) in elastic scattering of electrons on the spin-0 nuclei or the nuclei with equal numbers of neutrons and protons \cite{donn79}. However, it no longer holds when using other gauge models with $\beta_A^{(0)} \ne 0$. For the $^7$Li and $^7$Be nuclei, the results obtained depend on their multipole form factor in all gauge models because all the nuclear form factors have both isoscalar and isovector components.

\vspace{5pt}

\begin{figure}[htbp]
	\includegraphics[width=0.5\textwidth]{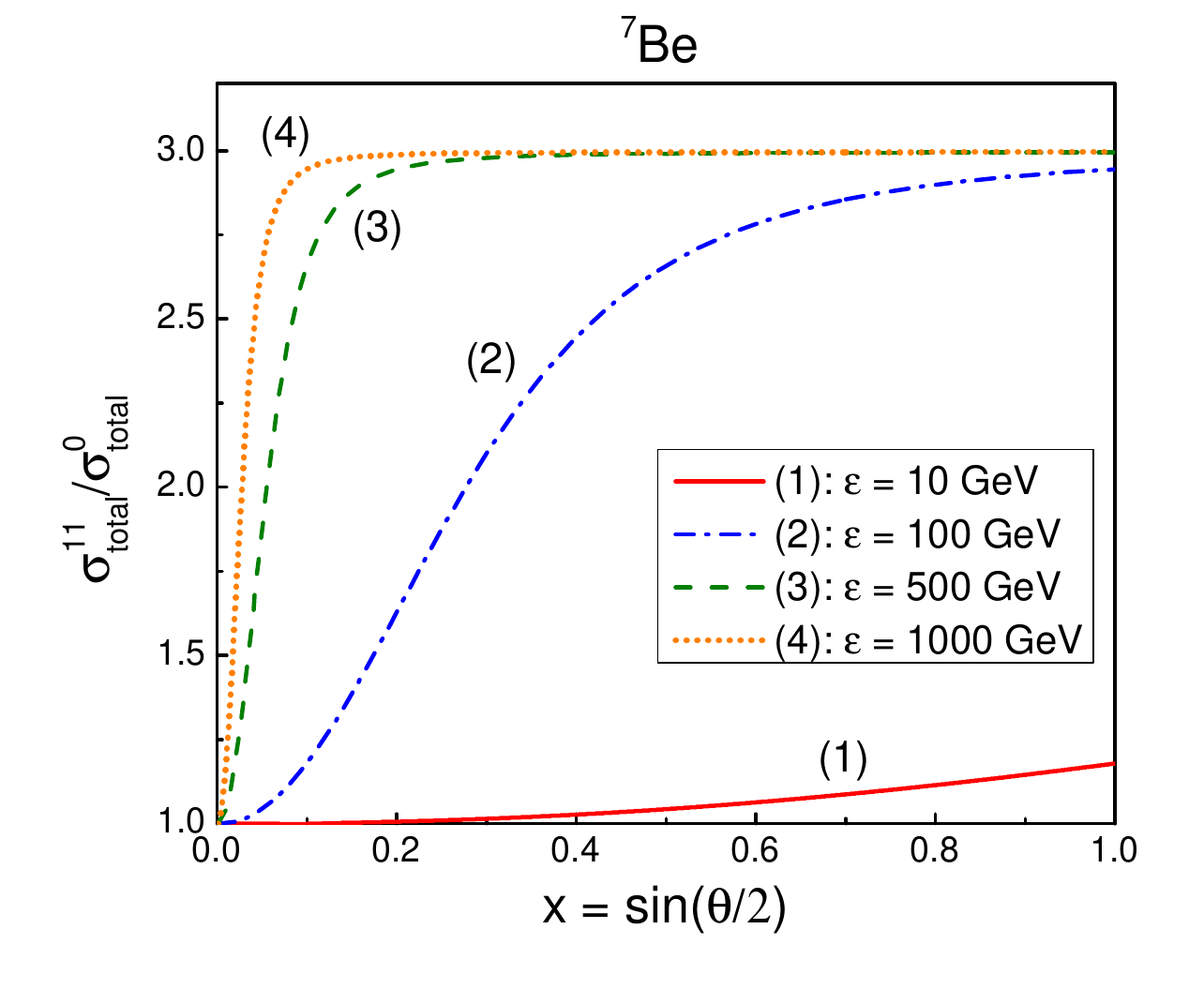}
	\vspace{-28pt}
	\caption{%	
		The polarized components $\sigma_{\text{total}}^{11}$ versus the
unpolarized ones $\sigma_{\text{total}}^{0}$ in the total polarized cross sections obtained
for $^{7}$Be at different electron energies $\varepsilon$ = 10, 100, 500, and 1000 GeV and scattering angles $\theta$.}
	\label{fig3}
\end{figure}

\vspace{5pt}
Note that the exponential functions in the numerator and denominator cancel each other out when calculating all considered ratios, while the scattering cross section itself rapidly decreases toward zero with increasing scattering angles at high energies, as previously reported in Refs. \cite{luon23,truo24}. Therefore, their plots remain fully visible across all scattering angles. Although the energy value of 1000 GeV belongs to the quark interaction region, which may no longer hold for electron-nucleus scattering in practice, we would like to show the ratios under consideration as in Figs. 1-3 and Table I to clearly depict their general patterns in the high-energy range. However, our data table only lists the ratios at narrow angles since their values at larger angles are not particularly meaningful once all the scattering cross sections are extremely small. 

\vspace {5pt}
In general, the results in Table I indicate that the information obtained in each scattering process strongly depends on how electron polarization changes. Specifically, once electron polarization remains unchanged, the scattering cross section $\sigma_{\text{total}}^{\text{pol.}}(1,1)$ is significantly larger than that obtained when the outgoing beam becomes unpolarized $\sigma_{\text{total}}^{\text{pol.}}(1,0)$, in particular at energies above 100 GeV and scattering angles $\theta = 5^{\circ} - 8^{\circ}$. Since the unpolarized term $\sigma_{\text{total}}^{0}$ in the total polarized cross section is identical in all cases, we can also make a direct comparison between $\sigma_{\text{total}}^{10}$ and $\sigma_{\text{total}}^{11}$. At energies below 10 GeV, the $\sigma_{\text{total}}^{10}/\sigma_{\text{total}}^{11}$ ratios are very small, with the highest order of magnitude being only around $10^{-4}$ at $\theta = 8^{\circ}$. However, at higher energy scales, these ratios increase remarkably, reaching up to several tens of percent. At the same energy and scattering angle, except for $\theta = 0^{\circ}$, their magnitudes follow an ascending order of $^{7}$Li, $^{6}$Li, and $^{7}$Be nuclei. To further clarify the impact of electron polarization in each scattering process, let us now consider the ratios $\sigma_{\text{total}}^{10}/\sigma_{\text{total}}^{0}$ and $\sigma_{\text{total}}^{11}/\sigma_{\text{total}}^{0}$. Similar to the $\sigma_{\text{total}}^{10}/\sigma_{\text{total}}^{11}$ ratio, the contribution of $\sigma_{\text{total}}^{10}$ can be neglected at 10 GeV, since its highest order of magnitude relative to $\sigma_{\text{total}}^{0}$ at $\theta = 8^{\circ}$ is only around $10^{-4}$. In contrast, $\sigma_{\text{total}}^{11}$ is nearly equal to $\sigma_{\text{total}}^{0}$ at energies not more than 10 GeV. At higher energy scales, the contribution of $\sigma_{\text{total}}^{10}$ becomes significant, whereas $\sigma_{\text{total}}^{11}$ even exceeds $\sigma_{\text{total}}^{0}$ by several times, as shown in Fig. 3 for  $^{7}$Be. 

\vspace {5pt}
The unpolarized and polarized scattering cross sections in cm$^2$/sr corresponding to several small scattering angles are also presented in Table II, considering the energy range that current electron accelerators can test. The results show that, at energies of several hundred GeV and above, the cross section approaches zero rapidly as the scattering angle increases. This implies that the electron could penetrate deeply inside the nucleus and nucleon, reaching the quark-size region and almost go straight after scattering, which is relevant to extremely small values of the cross section for nonzero scattering angles. The interaction between the electron and the nucleus is then more likely to be through smaller components, namely the quarks, rather than through its nucleons. In this sense, the scattering cross section ratios presented in Table I and Figs. 1-3 also provide important information that helps to clarify the role of electron polarization together with the weak interaction in the quark-size region, since the nucleon form factor in use decreases inversely with the square of the momentum transfer. 

\vspace{5pt}
Overall, all the ratios under consideration do not depend on the nuclear recoil factor and the center-of-mass motion. Furthermore, the calculations use the same harmonic oscillator parameter value for both polarized and unpolarized scattering, so that if we adjust the above parameters, the results are still the same. They show a little difference among selected nuclei in the scattering processes at energies below 10 GeV. At higher energies, their magnitudes increase in the order of $^{7}$Li, $^{6}$Li, and $^{7}$Be nuclei. Particularly, there are remarkable discrepancies in the contribution of electron polarization to the scattering cross section when comparing stable and unstable nuclei. The $^{7}$Li -$^{7}$Be pair proves this remark, although both have the same nucleon numbers, in which $^7$Be is an unstable nucleus with a half-life of around 53.22 days. In contrast, for the $^{6}$Li -$^{7}$Li stable pair, these discrepancies are much smaller, even though their nucleon numbers are slightly different. It can be explained that the weak interaction between electrons and nuclei strongly relates to nuclear stability. This property affects the correlation between the weak interaction and electron polarization, along with their contribution in each scattering process. Furthermore, although the $\sigma_{\text{total}}^{10}$ term can take negative values, its magnitude is significantly smaller than the $\sigma_{\text{total}}^{0}$ and $\sigma_{\text{total}}^{11}$ ones. Therefore, the total differential cross section also depends on the correlation between the electron spin orientations before and after scattering and remains positive in all cases. In particular, when the longitudinally polarized electrons move straight forward after scattering, the electron polarization-weak interaction correlation is not exhibited. This is evidenced by the ratios at $\theta = 0^{\circ}$ in Table I. Their values remain nearly unchanged at all considered energies, except for $\sigma_{\text{total}}^{10}$, whose contribution can be omitted. So, besides nuclear spin orientation, it is evident that electron polarization and its transformation under the weak interaction have a significant impact on electron-nucleus scattering processes at high energies. The scattering cross section ratios in the cases $\sigma_{\text{total}}^{\text{pol.}}(-1,0)$ and $\sigma_{\text{total}}^{\text{pol.}}(-1,-1)$ can be analyzed similarly and we can extend to other light nuclei with the multipole form factors being determined by using the formulas in Ref. \cite{truo24}. 
\vspace{10pt}

{\Large \textbf{4. Summary and Outlook}}

\vspace{5pt}
In conclusion, we have performed a theoretical study on the polarized electron scattering from unoriented light nuclei, utilizing the unified electroweak theory. Numerical calculations of the scattering cross sections in different electron polarization cases for three $^{6,7}$Li and $^{7}$Be light nuclei are obtained by directly calculating their multipole form factors. The results obtained are then compared and/or correlated with the corresponding cross sections for the unpolarized electron scattering case. It has been found that at the scattering angle of $\theta\simeq 0^\circ$, the longitudinal electron polarization does not explicitly correlate with the weak interaction for all energy scales. Such a correlation is strongly exhibited at the remaining scattering angles when the incoming electron energy passes 10 GeV. By comparing the results obtained for stable and unstable
nuclei, we found that the nuclear stability may significantly influence the contribution of electron polarization to the scattering cross section. 

\vspace {5pt}
This research direction can be extended in many ways, specifically considering various state transitions within a single nucleus or different nuclei. Then, the asymmetry resulting solely from electron polarization and the weak interaction can be figured out. In addition, other interesting results are expected once the nuclear spin orientation and charge-changing weak current are included. Moreover, we could apply the multipole expansion for the scattering cross section within the unified electroweak theory to directly study the EMC effect \cite{bemi23,gome94,seel09,schm19} in a target nucleus using both unpolarized and polarized electrons, e.g., through quasielastic electron scattering on $^7$Li and $^7$Be nuclei corresponding to a nuclear excitation from the ground state to the nearest state. More specifically, we shall calculate the cumulative cross section contributed by separate nucleons inside a certain nucleus and compare it with that made of a similar number of nucleons bound inside the same nucleus. This means that we can investigate the above effect on only a single nucleus instead of indirectly comparing its scattering cross section with that of a lighter nucleus, in which calculations are based on nucleons and nuclei instead of quarks. These extensive studies will provide further vital information, contributing to a thorough understanding of the nature of the EMC effect in inelastic electron-nucleus scattering. 

%\acknowledgments
\vspace{5pt}

M. T. Vo would like to thank Dr. Sc. Luong Zuyen Phu for his useful discussions on this work.
\vspace{0pt}
%%%%%%%%%%%%%%%%%%%%%%%%%%%%%%%%%%%%%%%%%%%%
\twocolumngrid

%%%%%%%%%%%%%%%%%%%%%%%%%%%%%%%%%%%%%%%%%%%%%%%%%

\onecolumngrid
\begin{table*}[htt]
	\caption{Calculated cross section ratios for $^{6,7}$Li and $^{7}$Be nuclei obtained within the unified electroweak theory at different electron
energies $\varepsilon$ = 10, 100, 500, and 1000 GeV and small scattering angles.}
	\centering
	\setlength{\tabcolsep}{3 pt}
	\renewcommand{\arraystretch}{1.0}
	\resizebox{\textwidth}{!}{%
		\begin{tabular}{>{\centering\arraybackslash} ccccccccccccccccc}
			\toprule
			\midrule[0.8pt]
			\multirow{2}{*}{\Large$\varepsilon$} & \multirow{2}{*}{\Large$\theta$} & \multicolumn{3}{c}{ \Large$ \sigma^{\text{pol.}}_{\text{total}}(1,0) / \sigma^{\text{unpol.}}_{\text{total}}$} & \multicolumn{3}{c}{\Large$\sigma^{\text{pol.}}_{\text{total}}(1,1) / \sigma^{\text{unpol.}}_{\text{total}}$} & \multicolumn{3}{c}{\Large$\sigma^{10}_{\text{total}} / \sigma^{11}_{\text{total}}$} & \multicolumn{3}{c}{\Large$\sigma^{10}_{\text{total}} / \sigma^{0}_{\text{total}}$} & \multicolumn{3}{c}{\Large$\sigma^{11}_{\text{total}} / \sigma^{0}_{\text{total}}$}  \\
			\cmidrule(lr){3-5} \cmidrule(lr){6-8} \cmidrule(lr){9-11} \cmidrule(lr){12-14} \cmidrule(lr){15-17}
			\large(GeV)& & {\large$^{7}$Li} & {\large$^{6}$Li} & {\large$^{7}$Be} & {\large$^{7}$Li} & {\large$^{6}$Li} & {\large$^{7}$Be} & {\large\large$^{7}$Li} & {\large$^{6}$Li} & {\large$^{7}$Be} & {\large$^{7}$Li} & {\large$^{6}$Li} & {\large$^{7}$Be} & {\large$^{7}$Li} & {\large$^{6}$Li} & {\large$^{7}$Be} \\
			\midrule
			\multirow{8}{*}{10} & $0^\circ$ & 0.5000 & 0.5000 & 0.5000 & 1.0000 & 1.0000 & 1.0000 & \makecell{9.8236 \\$\times 10^{-10}$} & \makecell{4.6965 \\$\times 10^{-10}$}  & \makecell{2.1329 \\$\times 10^{-10}$}  & \makecell{9.8236 \\$\times 10^{-10}$}  & \makecell{ 4.6965 \\ $\times 10^{-10}$ }  & \makecell{2.1329 \\$\times 10^{-10}$}  & 1.0000 & 1.0000 & 1.0000 \\ [3.0 ex]
			
			& $3^\circ$ & 0.5000 & 0.5000 & 0.5000 & 1.0000 & 1.0000 & 1.0000 & \makecell{2.1903 \\$\times 10^{-5}$} & \makecell{3.2235 \\$\times 10^{-5}$} & \makecell{4.8553 \\$\times 10^{-6}$} & \makecell{2.1903 \\$\times 10^{-5}$} & \makecell{3.2237 \\$\times 10^{-5}$} & \makecell{4.8554 \\$\times 10^{-6}$} & 1.0000 & 1.0001 & 1.0000 \\ [3.0 ex]
			& $5^\circ$ & 0.5000 & 0.5000 & 0.5000 & 0.9999 & 1.0001 & 1.0000 & \makecell{-5.9973 \\$\times 10^{-5}$} & \makecell{8.9254 \\$\times 10^{-5}$} & \makecell{-3.4234 \\$\times 10^{-5}$} & \makecell{-5.9966 \\$\times 10^{-5}$} & \makecell{8.9270 \\$\times 10^{-5}$} & \makecell{-3.4231 \\$\times 10^{-5}$} & 0.9999 & 1.0002 & 0.9999 \\ [3.0 ex]
			& $8^\circ$ & 0.4999 & 0.5001 & 0.4998 & 0.9998 & 1.0002 & 0.9996 & \makecell{-1.9758 \\$\times 10^{-4}$} & \makecell{2.2865 \\$\times 10^{-4}$} & \makecell{-3.5613 \\$\times 10^{-4}$} & \makecell{-1.9750 \\$\times 10^{-4}$} & \makecell{2.2876 \\$\times 10^{-4}$} & \makecell{-3.5588 \\$\times 10^{-4}$} & 0.9996 & 1.0005 & 0.9993 \\
			\midrule
			\multirow{8}{*}{100} & $0^\circ$ & 0.5000 & 0.5000 & 0.5000 & 1.0000 & 1.0000 & 1.0000 & \makecell{9.8194 \\$\times 10^{-8}$} & \makecell{4.6965 \\$\times 10^{-8}$} & \makecell{2.1317 \\$\times 10^{-8}$} & \makecell{9.8194 \\$\times 10^{-8}$} & \makecell{4.6965 \\$\times 10^{-8}$} & \makecell{2.1317 \\$\times 10^{-8}$} & 1.0000 & 1.0000 & 1.0000 \\ [3.0 ex]
			
			& $3^\circ$ & 0.5005 & 0.5016 & 0.5030 & 1.0010 & 1.0032 & 1.0059 & 0.0010 & 0.0032 & 0.0059 & 0.0010 & 0.0032 & 0.0059 & 1.0021 & 1.0064 & 1.0119 \\
			& $5^\circ$ & 0.5018 & 0.5044 & 0.5089 & 1.0036 & 1.0088 & 1.0179 & 0.0036 & 0.0087 & 0.0172 & 0.0036 & 0.0088 & 0.0179 & 1.0072 & 1.0177 & 1.0357 \\
			& $8^\circ$ & 0.5049 & 0.5112 & 0.5230 & 1.0097 & 1.0223 & 1.0461 & 0.0096 & 0.0214 & 0.0422 & 0.0097 & 0.0223 & 0.0461 & 1.0195 & 1.0447 & 1.0921 \\
			\midrule
			\multirow{8}{*}{500} & $0^\circ$ & 0.5000 & 0.5000 & 0.5000 & 1.0000 & 1.0000 & 1.0000 & \makecell{2.4294 \\$\times 10^{-6}$} & \makecell{1.1741 \\$\times 10^{-6}$} & \makecell{5.2571 \\$\times 10^{-7}$} & \makecell{2.4294 \\$\times 10^{-6}$} & \makecell{1.1741 \\$\times 10^{-6}$} & \makecell{5.2571 \\$\times 10^{-7}$} & 1.0000 & 1.0000 & 1.0000 \\ [3.0 ex]
			
			& $3^\circ$ & 0.5167 & 0.5371 & 0.5766 & 1.0334 & 1.0741 & 1.1533 & 0.0313 & 0.0645 & 0.1173 & 0.0334 & 0.0741 & 0.1533 & 1.0668 & 1.1482 & 1.3066 \\
			& $5^\circ$ & 0.5409 & 0.5894 & 0.6804 & 1.0818 & 1.1788 & 1.3609 & 0.0703 & 0.1317 & 0.2096 & 0.0818 & 0.1788 & 0.3609 & 1.1635 & 1.3577 & 1.7217 \\
			& $8^\circ$ & 0.5807 & 0.6722 & 0.8222 & 1.1614 & 1.3444 & 1.6444 & 0.1220 & 0.2039 & 0.2815 & 0.1614 & 0.3444 & 0.6444 & 1.3227 & 1.6889 & 2.2889 \\
			\midrule
			\multirow{8}{*}{1000} & $0^\circ$ & 0.5000 & 0.5000 & 0.5000 & 1.0000 & 1.0000 & 1.0000 & \makecell{9.3906 \\$\times 10^{-6}$} & \makecell{4.6964 \\$\times 10^{-6}$} & \makecell{2.0144 \\$\times 10^{-6}$} & \makecell{9.3908 \\$\times 10^{-6}$} & \makecell{4.6965 \\$\times 10^{-6}$} & \makecell{2.0144 \\$\times 10^{-6}$} & 1.0000 & 1.0000 & 1.0000 \\ [3.0 ex]
			
			& $3^\circ$ & 0.5545 & 0.6183 & 0.7335 & 1.1089 & 1.2365 & 1.4670 & 0.0894 & 0.1606 & 0.2415 & 0.1089 & 0.2365 & 0.4670 & 1.2179 & 1.4730 & 1.9340 \\
			& $5^\circ$ & 0.6037 & 0.7173 & 0.8839 & 1.2074 & 1.4346 & 1.7677 & 0.1466 & 0.2325 & 0.3028 & 0.2074 & 0.4346 & 0.7677 & 1.4148 & 1.8693 & 2.5355 \\
			& $8^\circ$ & 0.6498 & 0.7995 & 0.9635 & 1.2996 & 1.5989 & 1.9270 & 0.1873 & 0.2725 & 0.3248 & 0.2996 & 0.5989 & 0.9270 & 1.5992 & 2.1979 & 2.8541 \\
			\midrule[0.8pt]
			\bottomrule
	\end{tabular}}
\end{table*}

\onecolumngrid
\begin{table*}[htt]
	\caption{Numerically unpolarized and polarized electron elastic scattering cross sections (cm$^2$/sr) for $^{6,7}$Li and $^{7}$Be nuclei obtained within the unified electroweak theory at different electron
energies $\varepsilon$ = 10, 50, 100, and 120 GeV and small angles.}
	\centering
	\setlength{\tabcolsep}{8 pt}
	\renewcommand{\arraystretch}{1.0}
	\resizebox{\textwidth}{!}{%
		\begin{tabular}{>{\centering\arraybackslash}ccccccccccc}
			\toprule
			\midrule[0.8pt]
			\multirow{2}{*}{\large$\varepsilon$} & \multirow{2}{*}{\large$\theta$} & \multicolumn{3}{c}{\large$^{7}$Li} & \multicolumn{3}{c}{\large$^{6}$Li} & \multicolumn{3}{c}{\large$^{7}$Be} \\
			\cmidrule(lr){3-5} \cmidrule(lr){6-8} \cmidrule(lr){9-11} 
			(GeV)& & {$\sigma^{\text{unpol.}}_{\text{total}}$} & {$\sigma^{\text{pol.}}_{\text{total}}(1,0)$} & {$\sigma^{\text{pol.}}_{\text{total}}(1,1)$} & {$\sigma^{\text{unpol.}}_{\text{total}}$} & {$\sigma^{\text{pol.}}_{\text{total}}(1,0)$} & {$\sigma^{\text{pol.}}_{\text{total}}(1,1)$} & {$\sigma^{\text{unpol.}}_{\text{total}}$} & {$\sigma^{\text{pol.}}_{\text{total}}(1,0)$} & {$\sigma^{\text{pol.}}_{\text{total}}(1,1)$} \\
			\midrule
			\multirow{8}{*}{10} & $0^\circ$ & \makecell{5.1899 \\$\times 10^{-19}$} & \makecell{2.5949 \\$\times 10^{-19}$} & \makecell{5.1899 \\$\times 10^{-19}$} & \makecell{5.1898 \\$\times 10^{-19}$} & \makecell{2.5949 \\$\times 10^{-19}$} & \makecell{5.1898 \\$\times 10^{-19}$} & \makecell{2.0759 \\$\times 10^{-18}$} & \makecell{1.0379 \\$\times 10^{-18}$}  & \makecell{2.0759 \\$\times 10^{-18}$} \\ [2.5 ex]
            
			& $3^\circ$ & \makecell{1.0666 \\$\times 10^{-30}$} & \makecell{0.5333 \\$\times 10^{-30}$} & \makecell{1.0666 \\$\times 10^{-30}$} & \makecell{6.3607 \\$\times 10^{-34}$} & \makecell{3.1803 \\$\times 10^{-34}$} & \makecell{6.3607 \\$\times 10^{-34}$} & \makecell{3.3791 \\$\times 10^{-30}$} & \makecell{1.6896 \\$\times 10^{-30}$} & \makecell{3.3791 \\$\times 10^{-30}$} \\ [2.5 ex]
            
			& $5^\circ$ & \makecell{9.7062 \\$\times 10^{-33}$} & \makecell{4.8531 \\$\times 10^{-33}$} & \makecell{9.7052 \\$\times 10^{-33}$} & \makecell{1.0157 \\$\times 10^{-33}$} & \makecell{0.5079 \\$\times 10^{-33}$} & \makecell{1.0158 \\$\times 10^{-33}$} & \makecell{1.9498 \\$\times 10^{-32}$} & \makecell{0.9749 \\$\times 10^{-32}$} & \makecell{1.9498 \\$\times 10^{-32}$} \\ [2.5 ex]
            
			& $8^\circ$ & \makecell{1.8223 \\$\times 10^{-37}$} & \makecell{0.9110 \\$\times 10^{-37}$} & \makecell{1.8219 \\$\times 10^{-37}$} & \makecell{1.7298 \\$\times 10^{-39}$} & \makecell{0.8651 \\$\times 10^{-39}$} & \makecell{1.7301 \\$\times 10^{-39}$} & \makecell{9.6850 \\$\times 10^{-38}$} & \makecell{4.8406 \\$\times 10^{-38}$} & \makecell{9.6811 \\$\times 10^{-38}$} \\
			\midrule
			\multirow{8}{*}{50} & $0^\circ$ & \makecell{2.0717 \\$\times 10^{-20}$} & \makecell{1.0358 \\$\times 10^{-20}$} & \makecell{2.0717 \\$\times 10^{-20}$} & \makecell{2.0713 \\$\times 10^{-20}$} & \makecell{1.0357 \\$\times 10^{-20}$} & \makecell{2.0713 \\$\times 10^{-20}$} & \makecell{8.2870 \\$\times 10^{-20}$} & \makecell{4.1435 \\$\times 10^{-20}$} & \makecell{8.2870 \\$\times 10^{-20}$} \\ [2.5 ex]
			
			& $2^\circ$ & \makecell{2.3128 \\$\times 10^{-40}$} & \makecell{1.1562 \\$\times 10^{-40}$} & \makecell{2.3123 \\$\times 10^{-40}$} & \makecell{1.1994 \\$\times 10^{-42}$} & \makecell{0.6005 \\$\times 10^{-42}$} & \makecell{1.2011 \\$\times 10^{-42}$} & \makecell{ 5.8293 \\$\times 10^{-41}$} & \makecell{2.9129 \\$\times 10^{-41}$} & \makecell{5.8252 \\$\times 10^{-41}$} \\ [2.5 ex]
            
			& $3^\circ$ & \makecell{9.0803 \\$\times 10^{-55}$} & \makecell{4.5401 \\$\times 10^{-55}$} & \makecell{9.0803 \\$\times 10^{-55}$} & \makecell{3.7204 \\$\times 10^{-58}$} & \makecell{1.8662 \\$\times 10^{-58}$} & \makecell{3.7323 \\$\times 10^{-58}$} & \makecell{1.4712 \\$\times 10^{-55}$} & \makecell{0.7357 \\$\times 10^{-55}$} & \makecell{1.4713 \\$\times 10^{-55}$} \\ [2.5 ex]
            
			& $5^\circ$ & \makecell{5.6798 \\$\times 10^{-100}$} & \makecell{2.8416 \\$\times 10^{-100}$} & \makecell{5.6832 \\$\times 10^{-100}$} & \makecell{1.0340 \\$\times 10^{-107}$} & \makecell{0.5215 \\$\times 10^{-107}$} & \makecell{1.0431 \\$\times 10^{-107}$} & \makecell{1.1469 \\$\times 10^{-100}$} & \makecell{0.5756 \\$\times 10^{-100}$} & \makecell{1.1513 \\$\times 10^{-100}$} \\
			\midrule
			\multirow{8}{*}{100} & $0^\circ$ & \makecell{5.1464 \\$\times 10^{-21}$} & \makecell{2.5732 \\$\times 10^{-21}$} & \makecell{5.1464 \\$\times 10^{-21}$} & \makecell{5.1422 \\$\times 10^{-21}$} & \makecell{2.5711 \\$\times 10^{-21}$} & \makecell{5.1422 \\$\times 10^{-21}$} & \makecell{2.0587 \\$\times 10^{-20}$} & \makecell{1.0294 \\$\times 10^{-20}$} & \makecell{2.0587 \\$\times 10^{-20}$}  \\ [2.5 ex]
			
			& $1^\circ$ & \makecell{1.2618 \\$\times 10^{-39}$} & \makecell{0.6308 \\$\times 10^{-39}$} & \makecell{1.2615 \\$\times 10^{-39}$} & \makecell{6.6551 \\$\times 10^{-42}$} & \makecell{3.3276 \\$\times 10^{-42}$} & \makecell{6.6551 \\$\times 10^{-42}$} & \makecell{3.2308 \\$\times 10^{-40}$} & \makecell{1.6144 \\$\times 10^{-40}$} & \makecell{3.2285 \\$\times 10^{-40}$} \\ [2.5 ex]
            
			& $2^\circ$ & \makecell{2.7431 \\$\times 10^{-74}$} & \makecell{1.3718 \\$\times 10^{-74}$} & \makecell{2.7436 \\$\times 10^{-74}$} & \makecell{1.4419 \\$\times 10^{-79}$} & \makecell{0.7459 \\$\times 10^{-79}$} & \makecell{1.4918 \\$\times 10^{-79}$} & \makecell{5.0209 \\$\times 10^{-75}$} & \makecell{2.5150 \\$\times 10^{-75}$} & \makecell{5.0304 \\$\times 10^{-75}$} \\ [2.5 ex]
            
			& $3^\circ$ & \makecell{1.4273 \\$\times 10^{-130}$} & \makecell{0.7144 \\$\times 10^{-130}$} & \makecell{1.4287 \\$\times 10^{-130}$} & \makecell{2.0478 \\$\times 10^{-141}$} & \makecell{1.0999 \\$\times 10^{-141}$} & \makecell{2.1995 \\$\times 10^{-141}$} & \makecell{3.0663 \\$\times 10^{-131}$} & \makecell{1.5423 \\$\times 10^{-131}$} & \makecell{3.0844 \\$\times 10^{-131}$} \\
			\midrule
			\multirow{8}{*}{120} & $0^\circ$ & \makecell{3.5605 \\$\times 10^{-21}$} & \makecell{1.7803 \\$\times 10^{-21}$} & \makecell{3.5605 \\$\times 10^{-21}$} & \makecell{3.5564 \\$\times 10^{-21}$} & \makecell{1.7782 \\$\times 10^{-21}$} & \makecell{3.5564 \\$\times 10^{-21}$} & \makecell{1.4243 \\$\times 10^{-20}$} & \makecell{0.7121 \\$\times 10^{-20}$} & \makecell{1.4243 \\$\times 10^{-20}$}  \\ [2.5 ex]
			
			& $1^\circ$ & \makecell{1.5070 \\$\times 10^{-44}$} & \makecell{0.7533 \\$\times 10^{-44}$} & \makecell{1.5067 \\$\times 10^{-44}$} & \makecell{3.7649 \\$\times 10^{-47}$} & \makecell{1.8825 \\$\times 10^{-47}$} & \makecell{3.7649 \\$\times 10^{-47}$} & \makecell{2.6684 \\$\times 10^{-45}$} & \makecell{1.3334 \\$\times 10^{-45}$} & \makecell{2.6665 \\$\times 10^{-45}$} \\  [2.5 ex]
            
			& $2^\circ$ & \makecell{3.9184 \\$\times 10^{-94}$} & \makecell{1.9604 \\$\times 10^{-94}$} & \makecell{3.9204 \\$\times 10^{-94}$} & \makecell{2.2680 \\$\times 10^{-101}$} & \makecell{1.2748 \\$\times 10^{-101}$} & \makecell{2.5497 \\$\times 10^{-101}$} & \makecell{7.7925 \\$\times 10^{-95}$} & \makecell{3.9095 \\$\times 10^{-95}$} & \makecell{7.8190 \\$\times 10^{-95}$} \\ [2.5 ex]
            
			& $3^\circ$ & \makecell{6.4150 \\$\times 10^{-175}$} & \makecell{3.2126 \\$\times 10^{-175}$} & \makecell{6.4259 \\$\times 10^{-175}$} & \makecell{3.2141 \\$\times 10^{-190}$} & \makecell{1.9873 \\$\times 10^{-190}$} & \makecell{3.9742 \\$\times 10^{-190}$} & \makecell{1.4418 \\$\times 10^{-175}$} & \makecell{0.7274 \\$\times 10^{-175}$} & \makecell{1.4548 \\$\times 10^{-175}$} \\
			\midrule[0.8pt]
			\bottomrule
	    \end{tabular}}
\end{table*}

\end{document}